# Green low-cost carbon nanodots-polyurethane composites with novel anisotropic anti-quenching mechanism for strain sensing

*Yayuan Tian, Yan Zhao, Fengwen Kang, Fucong Lyu, Zebiao Li, Jian Lu\*, and Yang Yang Li\**

Y. Tian, Z. Li, Prof. Y. Y. Li
Center of Super-Diamond and Advanced Films (COSDAF), City University of Hong Kong, 83 Tat Chee Avenue, Kowloon, Hong Kong
E-mail: yangli@cityu.edu.hk

Y. Tian, Dr. F. Kang, Z. Li, Prof. J. Lu, Prof. Y. Y. Li,
Department of Materials Science and Engineering, City University of Hong Kong, 83 Tat Chee Avenue, Kowloon, Hong Kong
E-mail: jianlu@cityu.edu.hk

Dr. Y. Zhao, Dr. F. Kang, F. Lyu, Prof. J. Lu, Prof. Y. Y. Li
Department of Mechanical Engineering, City University of Hong Kong, 83 Tat Chee Avenue, Kowloon, Hong Kong

Prof. J. Lu, Prof. Y. Y. Li
Hong Kong Branch of National Precious Metals Material Engineering Research Centre, City University of Hong Kong, 83 Tat Chee Avenue, Kowloon, Hong Kong

Prof. J. Lu
Centre for Advanced Structural Materials, City University of Hong Kong Shenzhen Research Institute, Yuexing 1st Road, Shenzhen Hi-Tech Industrial Park, Nanshan District, Shenzhen, China

**Abstract**

A new type of nontoxic low-cost sensor is reported here, whose photoluminescence (PL) intensity is instantly responsive to the external strain applied over a large range (up to 250% strain). Highly stretchable fluorescent composites of carbon dots (CDs) and polyurethane (PU) are fabricated via a scalable green chemistry method by conveniently dispersing CDs in the aqueous solution of PU. It is discovered that, upon tensile deformation, the PL peak of the CD-PU film remains non-shifted but displays varied intensity. The observed PL responses to strain are ascribed to the enlarged inter-particle distance of CDs along the tensile direction (z axis), although a higher degree of aggregation is resulted in the other two axes. The PL-dependence on the anisotropic patterns of CDs in solid state points to a new mechanism to overcome aggregation-induced quenching by controlling the distribution behaviors of the fluorescent species.

**Introduction**

Fluorescent quantum dots are very attractive for many applications including anti-counterfeiting,[1] information protection,[2] displays and lighting,[3] but their development is greatly limited by either toxicity from heavy-metal elements such as arsenic (As) in GaAs, or aggregation-induced luminescence quenching. Carbon dots (CDs) as a new class of luminescent material have attracted much attention due to merits such as low toxicity, chemical stability, biocompatibility, and good dispersibility in water.[4] However, how to achieve fluorescence remains a major challenge for the solid-state CDs. To fulfill this task, the self-quenching effect, caused by the excessive resonance energy transfer (RET) or direct π-π interactions (similar to the organic fluorescent molecules), needs to be prohibited among CDs.[5] Current solutions rely on dispersing CDs into other matrixes, such as polymers (e.g., polyvinyl alcohol (PVA)),[6] organosilane,[7] and starch,[8] through either covalent bonding (e.g., with the silica network) or electrostatic interaction, so that the CDs can be anchored separately in the host to resist aggregation-induced self-quenching. Notably, although the impact of CD concentration in the host material is well documented, the influence of CD distribution pattern on their PL properties is rarely noticed.

Here we report a novel type of strain sensor based on CD-polyurethane composites. Interestingly, the PL properties of the CD-PU composites are highly responsive to strain applied, noticeable to the naked eye, indicating the aggregation-induced self-quenching can be overcome by mechanical stretching. Besides the large stretchability (*cf.* PVA), polyurethane (PU) is selected here as the host material also for their other merits of water-solubility, high biocompatibility, and good adhesion to various substrate materials, beneficial for their practical applications. To the best of our knowledge, the key impact of CD distribution pattern on their PL performance is reported for the first time. During the straining operation, the PU molecules were reoriented resulting in more separated CDs along the tensile axis, though a denser aggregation of CD is resulted in the other two axes. Notably, previous research on aggregation-induced quenching mainly focused on the isotropic CD systems while generally neglected the anisotropic ones. This discovery indicates that, in addition to concentration, another previously overlooked key factor for regulating PL is the CD distribution in the host.

In this work, a novel type of strain sensor is conveniently fabricated by ultrasonically dispersing a small amount of water-soluble CDs in an aqueous solution of PU, producing an elastic CD-PU composite film. The CD-PU strain sensor displays high detection range (e.g., from 0 to 250% strain), and elastic deformation recovery at strain 40% to realize the reusability (over 10 times). This strain sensor can be integrated/attached to wide-ranging substrate systems, allowing usage for safe and facile visualized strain sensing in different fields, such as for monitoring architecture health,[9] earthquake,[10] fatigue.[11] The current strain sensing technology is mainly based on mechano-induced optical change, and strain-induced change of electrical resistance or capacitance. The latter utilizes conductive nanomaterial composites (e.g., noble metal wires[12, 13] or carbon nanomaterials[14, 15]), generally needs delicate design, and displays narrow sensing ranges. Optical strain sensors possess the advantage of easy visualization, mainly employing mechanochromism,[16, 17] mechanoluminescence,[18] or photoluminescence.[19, 20] Mechanochromic strain sensors possess low stretchability and poor reusability, limiting their practical applications, while mechanoluminescence strain sensor is suited only for dynamic strain sensing rather than static strain sensing. By contrast, the CD-PU sensor reported here is conveniently fabricated from common low-cost chemicals using a green aqueous solution-based method, exhibiting fast response, great biocompatibility, and is suitable for dynamic and static strain measurements in various environments.

**Results and discussion**

*Physicochemical properties of carbon dots*

Carbon dots (CDs) were prepared by solvent-free thermal synthesis, with citric acid as carbon source and urea as nitrogen source heated at 170 °C in a mass ratio of 1:1. The morphology of the as-obtained carbon dots were examined using TEM **(Figure 1).** The as-prepared carbon dots display quasi-spherical nanoparticles with an average size of 3.2 nm (Figure 1a), with both amorphous and graphitic crystalline structure (Figure 1b). The well-resolved lattice fringes of graphitic crystalline structure show a spacing of 0.21 nm (Figure 1b), corresponding to the (100) in-plane lattice of graphene.[22] In addition, the CD's XRD pattern (Figure 1c) displays a sole diffraction peak at 2θ=27°, assigned to (002) diffraction pattern of graphitic carbon. Raman spectrum (Figure S1) also demonstrates its high graphitization, where the ordered G band at 1583 cm$^{-1}$ is stronger than the disordered D band at 1355 cm$^{-1}$, with G to D intensity ratio of 1.12.

The Fourier Transform Infrared (FTIR) spectrum (Figure 1d) revealed that the fabricated CDs possessed complex surface function groups, including hydrophilic groups such as O-H (3432 cm$^{-1}$), N-H (3200 cm$^{-1}$) and COOH (1709 cm$^{-1}$) leading to good solubility in water. The absorption band observed at 1601 cm$^{-1}$ are associated with the stretching vibration of C=C,[23] and strong absorbance at 1400 cm$^{-1}$ is due to the stretching vibration of C-N. Further FTIR study (Figure S2) showed no significant change of PU after CD incorporation, indicating no covalent or strong hydrogen bonding between CDs and PU.

XPS survey spectrum (**Figure 2**a) indicated that the fabricated CDs were mainly composed of carbon, nitrogen and oxygen. The high-resolution spectrum of C 1s (Figure 2b) was deconvoluted into four peaks, corresponding to $sp^2$ graphitic structure (C=C, 284.6 eV), $sp^3$ carbons (C-N, 285.5 eV) and (C-O, 286.4 eV), and carbonyl carbons (C=O, 288.2 eV). The high-resolution spectra of N 1s (Figure 2c) revealed the presence of pyridinic nitrogen (399.3 eV), pyrrolic nitrogen (400.1 eV) and graphitic nitrogen (401.2 eV), while the presence of C=O (531.4 eV) and C-O (532.7 eV) were observed in the high-resolution spectra of O 1s (Figure 2d). Therefore, the FTIR and XPS results confirmed that the CDs were functionalized with multiple surface groups and doped with nitrogen.

*PL properties*

The UV-Vis absorption spectrum of CDs in aqueous solution 0.1 mg/ml (i.e., 0.01 wt%) (**Figure 3**a) exhibited weak absorption peak at 274 nm and two obvious absorption peaks at 334 and 398 nm. The peak at 274 nm is assigned to π→π* transitions of aromatic $sp^2$ domains, whereas the peaks at 334 nm and 398 nm are attributed to n→π* transitions of C=O/C=N and surface functional groups, respectively. The PL emission spectra of CD aqueous solution exhibited excitation-dependent fluorescence with emission peaks ranging from 485 nm to 525 nm. The strongest fluorescence emission was detected at 525 nm under 420 nm excitation.

Solid-state luminescent CD-based composites have received increasing attention. Common methods for constructing solid-state CD-based composites usually employ substances such as

poly(vinyl alcohol) (PVA)[24] and silica.[25] Here, a convenient method for incorporating CDs into waterborne polyurethane is reported, forming solid-state composites possessing interesting strain-responsive fluorescence. Similar to CDs in aqueous solution, the CD-PU composite film (e.g., the one with CDs of 0.01 wt%) exhibits absorption peaks at 278 and 396 nm (Figure 3b), indicating the absorption mainly resulted from CDs in the composite. Again similarly, the PL emission spectra of the CD-PU film shows excitation-dependent fluorescence at different excitation wavelength, but narrower emission peaks from 494 nm to 522 nm than CDs in water, possibly due to the passivation of the CD surface functional groups by the surrounded PU network.

The effect of CD concentration in the composite film on the film's absorption and fluorescence properties was investigated. Notably, the PU host displayed quite featureless absorption with nearly nil absorption above 400 nm (Figure 3c). The absorption peaks of 0.01 wt% and 0.024 wt% CD-PU composite film were located at the same wavelengths (Figure 3c). Further addition of CDs in the PU host renders the absorption peak shifted from 278 nm to 297 nm which corresponds to π→π* transitions of aromatic $sp^2$ domains, induced by the self-assembled aggregation of CDs. Featuring π-conjugated molecular structure, CDs tend to aggregate with a higher concentration in PU. On the other hand, their surface groups at the zigzag edges restrain the π-π intermolecular interactions, favoring self-assembled packing with the head-to-tail molecular arrangement (J-type aggregation).[26, 27]

In good agreement with previous reports on CD-based composites,[28] the PL intensity of the solid-state CD composite films and the CD aqueous solutions varied as a function of the CD concentration (**Figure 4**). The pure PU film displays nil fluorescence under UV excitation of 412 nm. With a higher

CD concentration, the PL intensity surged to maximum with the CD concentrations around 0.024 to 0.048 wt%, dropped abruptly between 0.048 and 0.28 wt%, and then approached zero with further increased CD concentration, owning to the fluorescence quenching of CDs at high concentrations. The quantum yields of CD-PU films (Figure S3) reached a maximum value of 25% at 0.01 wt% CD, which gradually decreased with increased CD concentration, reaching 0.04% at 0.28 wt% CD. Interestingly, the appearance of the CD-PU films of different CD concentration under natural light display similar trend of transparency: the films were fairly transparent and colorless with CD concentration no higher than 0.048 wt%, which turned with increased CD concentration into dark brown to opaquely black (e.g., at 0.28 wt%), consistent with the observation previously reported for the CD-PVA film.[28] Noticeably, the concurrent abrupt change of PL and transparency at the same CD concentrations observed for the CD-PU films indicates that with increased concentration the agglomeration of CD in PU causes both severe self-quenching and high absorption effects.

The CD aqueous solution displayed similar PL behaviors with the CD-PU film (Figure 4), although the PL maximum appeared at a lower CD concentration (0.01wt%) than the latter (0.048 wt%) (Figure 4), possibly due to the more efficient non-radiative recombination in the water solution which aggravated the self-quenching effect. CDs tend to cluster while PU solid state host can anchor CDs in the host preventing CDs aggregation. This hypothesis was supported by the observation that, at the same CD concentration of 0.01wt%, the QY was merely 6.4% for the aqueous solution but 25% for the CD-PU composite (Figure S3).

Knowing that the CD concentration greatly affected the PL intensity of the CD-PU film, it is important to clarify that, for the strain sensors studied in this work, the overall CD concentration in the PU film was not significantly altered upon stretching (the Poisson's ratio of water-based PU film is about 0.4,[29, 30] and according to the literature,[31] the volume increase of the semicrystalline polymer is at most 6%), where it was rather the CD distribution pattern in the film that was changed.

*Strain sensing performance*

The CD-PU films with a relatively high CD concentration (e.g., 0.28%) exhibit nearly nil fluorescence. The stress loading was conducted on custom-made uniaxial loading machine (Figure S4). Interestingly, the PL intensity was significantly enhanced upon stretching the film over a large range, (e.g., 0 to 250% strain) (**Figure 5**a, Video in SI). At 0% strain, for the film containing 0.28 wt% CDs, the PL spectrum displayed a platform centered at 510 nm when excited at 405 nm. Upon uniaxial tensile loading at a strain rate of 22% $L_0$/s, an increase in PL intensity was observed (Figure 5a), with a steeper surge within the strain range of 0 to 50%, followed by a more gradual increase at higher strains (Figure 5b), reaching 1.4 times of the original intensity at 250% strain.

The optical responses of PU films with other CD concentrations were further studied (Figure 5d). The CD-PU films of 0.56 and 0.76 wt% CD displayed 1.6 and 1.8 times PL enhancement at 250% strain, respectively. Interestingly, different from the films of 0.28 or 0.56 wt% CDs which displayed a unidirectional PL increase till the maximum strain applied, PL of the film of 0.76 wt% CDs increased more dramatically at the initial stretching stage reaching 2.2 times at 100% strain, but then gradually dropped to 1.8 times at 250 % strain. This discrepancy observed for CD-PU with a higher CD concentration of 0.76 wt% can be ascribed to the fact that CDs are more likely to form aggregated particles/clusters in these composites.

It is noted that the CD-PU film recovered its original shape within the elastic strain range of 40%, in consistence with the literature,[32] accompanied by highly reversible PL responses. The cyclic uniaxial straining test of the CD-PU films approaching the elastic limit demonstrated their remarkable reliability and reusability (Figure 5d), holding high potential for practical applications.

In order to monitor strain distribution, a CD-PU film with a notch was uniaxially stretched to 125% strain under the UV irradiation of 365 nm (see the movie in the Supporting Information). Interestingly, more prominent PL enhancement was observed at the notch tip where the strain was more concentrated, indicating the capability of using the sensors in this study for visual detection of anisotropic strain distribution.

The strain sensing characteristics of the CD-PU composites reported here is compared with other reported strain sensors (**Table 1**). It is worth noting that the current approach offers a simple, green, low-cost, and scalable method for fabricating strain sensors suitable for detection of large strains (up to 250%), making use of the novel mechanism of overcoming the quenching effect of CDs in the solid-state PU host by uniaxial stretching.

*Strain sensing mechanism study*

To understand the PL behavior, it is necessary to point out that there are two factors that influence the PL intensity: one being the inter-particle distance that controls the self-quenching effect; the other being the total amount of CDs in the excited volume (denoted as *n*).

Let us first examine the influence of the inter-particle distance in the composite systems (**Figure 6**). It was clearly observed that the stretched film was elongated along the tensile direction, i.e., the z direction, and shrunk in the other two dimensions, i.e., the x and y directions. Notably, the overall volume of the CD-PU film was highly preservative, as the Poisson's ratio of PU remains very close to 0.5 over a wide strain range.[33] The observed PL enhancement for films with 0.28, 0.56, 0.76 wt% CDs discussed above is mainly attributed to the enlarged inter-particle distances over the spacing threshold in the z direction to forbid resonance energy transfer. In other words, the CDs are "diluted" in 1D in the stretched films to prevent PL quenching (Figure 6a). By contrast, previous research on controlling aggregation-induced quenching mainly focuses on dilution of the isotropic systems such as liquid solutions (Figure 6c), but largely neglects materials possessing anisotropic distribution of the fluorescent species, thus paying little attention on applying uniaxial tension to overcome the quenching effect.

We would like to further emphasize that the working mechanisms for overcoming the self-quenching effect induced by aggregation of the fluorescent species are fundamentally different for the anisotropic solid systems (e.g., stretched CD-PU films) and isotropic solution systems (e.g., colloidal solutions of CDs[34]). For liquid solution systems, dilution leads to PL increase through expanding the distance between the adjacent CDs homogeneously in three dimensions (Figure 6c). By contrast, the anti-quenching effect in the stretched CD-PU films is mainly due to uniaxially elongated distance between the CDs that surpasses the threshold for resonance energy transfer (Figure 6a).

With this anisotropic "dilution" mechanism proposed, one may wonder the PL response upon stretching of the CD-PU films with CD concentrations so low that their inter-particle distance is close to or beyond the quenching threshold. To answer this question, films of lower CD concentrations (e.g., 0.01, 0.062, or 0.096 wt%) were tested, unveiling PL behaviors very different from those higher concentration films: only slight decrease of PL was observed at tensile strains (Figure S6). To understand this difference, besides the impact of the inter-particle distance, the other factor, $n$, the total amount of CDs in the excited volume, is considered:

$$n = V \cdot C = A \cdot t \cdot C \tag{1}$$

where $V$ is the excited volume, $C$ the CD concentration of the PU film, $A$ the illuminated area, and $t$ the illuminated thickness. Considering that the film volume was almost unvaried, $C$ can be regarded as a constant upon stretching. Meanwhile, for the setup used in this study, the illuminated area or the light spot, $A$, was kept constant during the test. Hence, $n$ is mainly determined by the film thickness or the illumination penetration depth, whichever being thinner.

Notably, the PU film was significantly thinned upon stretching. Nevertheless, this thinning effect does not much affect the PL significantly for films of high CD concentration (e.g., 0.28 wt% or above), for these films are so absorptive (as evidenced by their very dark opaque appearances, see the photographs in Figure 4 insets) that the illumination can penetrate to only a shallow depth. Therefore, $n$ remain a constant for these films of highly concentrated CDs, whose PL will thus be mainly controlled by the inter-particle distances. On the other hand, for less absorptive or more translucent films of lower CD concentrations (e.g., 0.096% or below), film thinning upon stretching may significantly lower the amount of irradiated CDs. For example, the reflective excitation light was measured for films with 0.01 wt% CD of different thickness (0.2, 0.3, and 0.5 mm) (Figure. S5), displaying an apparent decrease of intensity for thinner films, indicating a higher transmission through the films at lower film thickness. As a result, for the films of more diluted CDs, $n$ is together with the inter-particle distance for determining the PL behaviors. Therefore, the PL decrease recorded for films of lower CD concentrations (Figure. S6) is attributed to the interplay of $n$ and the inter-particle distance.

**Conclusion**

By dispersing a small amount of carbon dots in water-soluble polyurethane, we have developed a new type of strain sensor whose photoluminescence properties are highly responsive to the applied strain over a wide range (up to 250%), broadening the application horizon of carbon dots into stretchable/wearable environmental-sensitive devices. The novel strain senor is cost effective, highly biocompatible, corrosion resistant, mass-producible with great fabrication convenience of green chemistry, and can be integrated/attached to wide-ranging substrate systems, holding great promise for safe and disposable visualized strain sensing of low cost in different fields, e.g., the automotive and aerospace industries. More importantly, investigation of the strain sensitivity unveils the distribution pattern-induced PL quenching/enhancement effect of carbon dots, indicating a new PL-controlling mechanism that may lay the foundation for many future discoveries and applications.

**Experimental**

*Synthesis of carbon dots*

CDs were fabricated by grounding citric acid (0.35 g) and urea (0.35 g) to fine powder in a mortar.

The mixture was transferred into a crucible and heated at 170 °C in a ventilated oven for 1 hr, and then cooled in air.[21] The as-prepared brown solid was dissolved in 170 ml of water, followed by centrifugation at 10,000 rpm for 15 min. The supernatant was collected and dialyzed against DI water through a dialysis membrane (MWCO 500-1000 Da) for 24 hr, and freeze-dried to obtain the final product of carbon dots.

*Construction of carbon dot-polyurethane composite film*

CDs were weighed and added into the aqueous solution of aliphatic polyurethane (Wanhua Chemistry), and ultrasonicated for 30 min. The dispersion solution was dropped into a glass container and cured for two days at ambient temperature, with final film thickness kept ~ 0.5 mm.

*Characterizations*

Transmission electron microscopy (TEM) observations were conducted on JEM-2100F FEG instrument operated at 200 kV. Powder X-ray diffraction (XRD) patterns were collected on a Bruker AXS D2 Phaser X-ray diffractometer with Cu−K$\alpha$ radiation ($\lambda$ = 1.5405 Å). Photoluminescence spectra were measured on a Hitachi F-4600 spectrophotometer, and quantum yield was measured on FLS 980. Fourier transform infrared (FTIR) spectra were obtained on a PE Spectrum 100. UV-Vis absorption spectra were measured on UV-VD (Agient 8453) for solution sample, and on UV-VIS-NIR Spectrophotometer (Shimadzu SolidSpec-3700/3700DUV) for solid sample. Raman spectra were measured on a Witec alpha300R Raman microscope with excitation wavelength of 532 nm.

For strain sensing experiment, stress loading was conducted on custom-made uniaxial loading machine, with the PL spectra simultaneously recorded using the Ocean Optics USB2000+ spectrometer connected to an optical fiber probe with numerical aperture (NA) 0.22, under the excitation light of 405 nm wavelength, that is by UV laser pointer 405 nm.

**Supporting Information**

Supporting Information is available from the Wiley Online Library or from the author.

**Acknowledgements**

This work was jointly supported by the Innovation and Technology Commission of HKSAR through Hong Kong Branch of National Precious Metals Material Engineering Research Center, and the City University of Hong Kong (Projects 7004643 and 7005077).

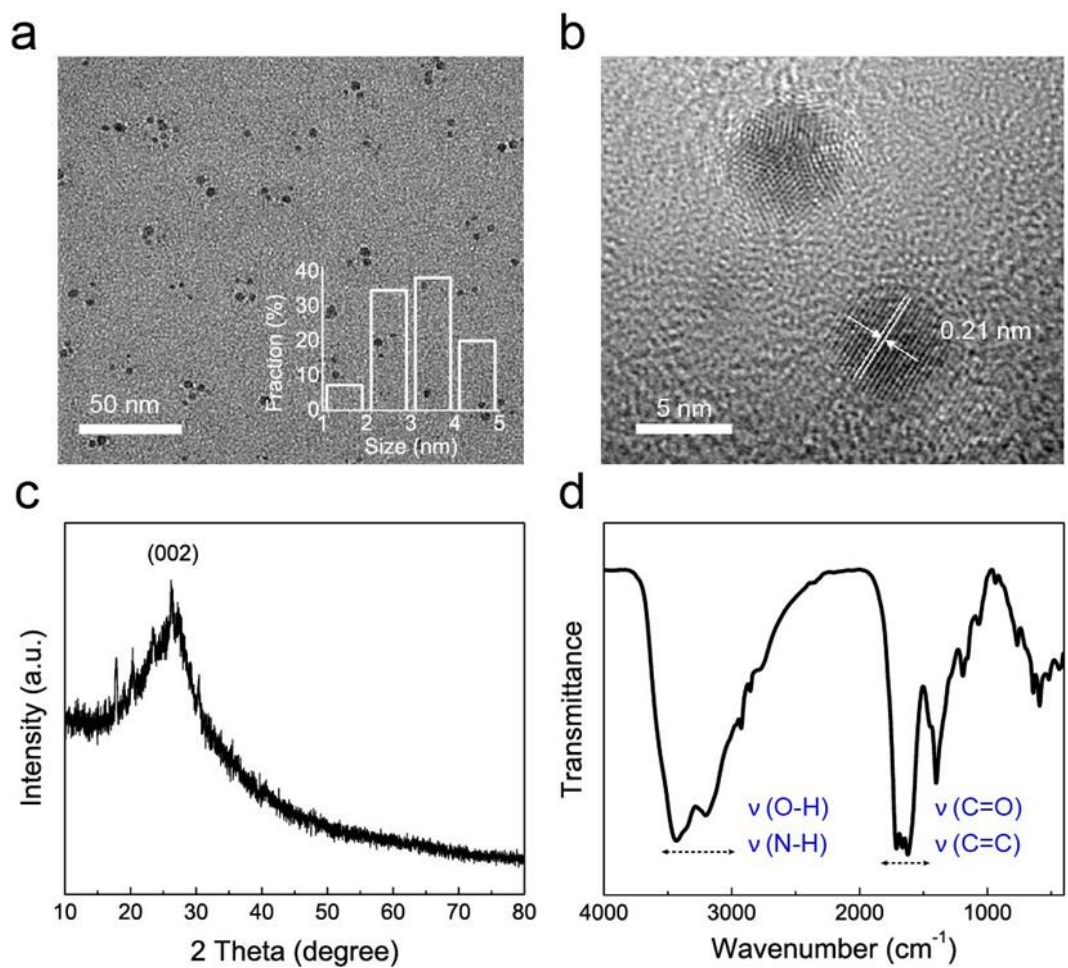

**Figure 1.** TEM image (a), HRTEM image (b), XRD pattern (c) and FTIR spectrum (d) of the fabricated carbon dots.

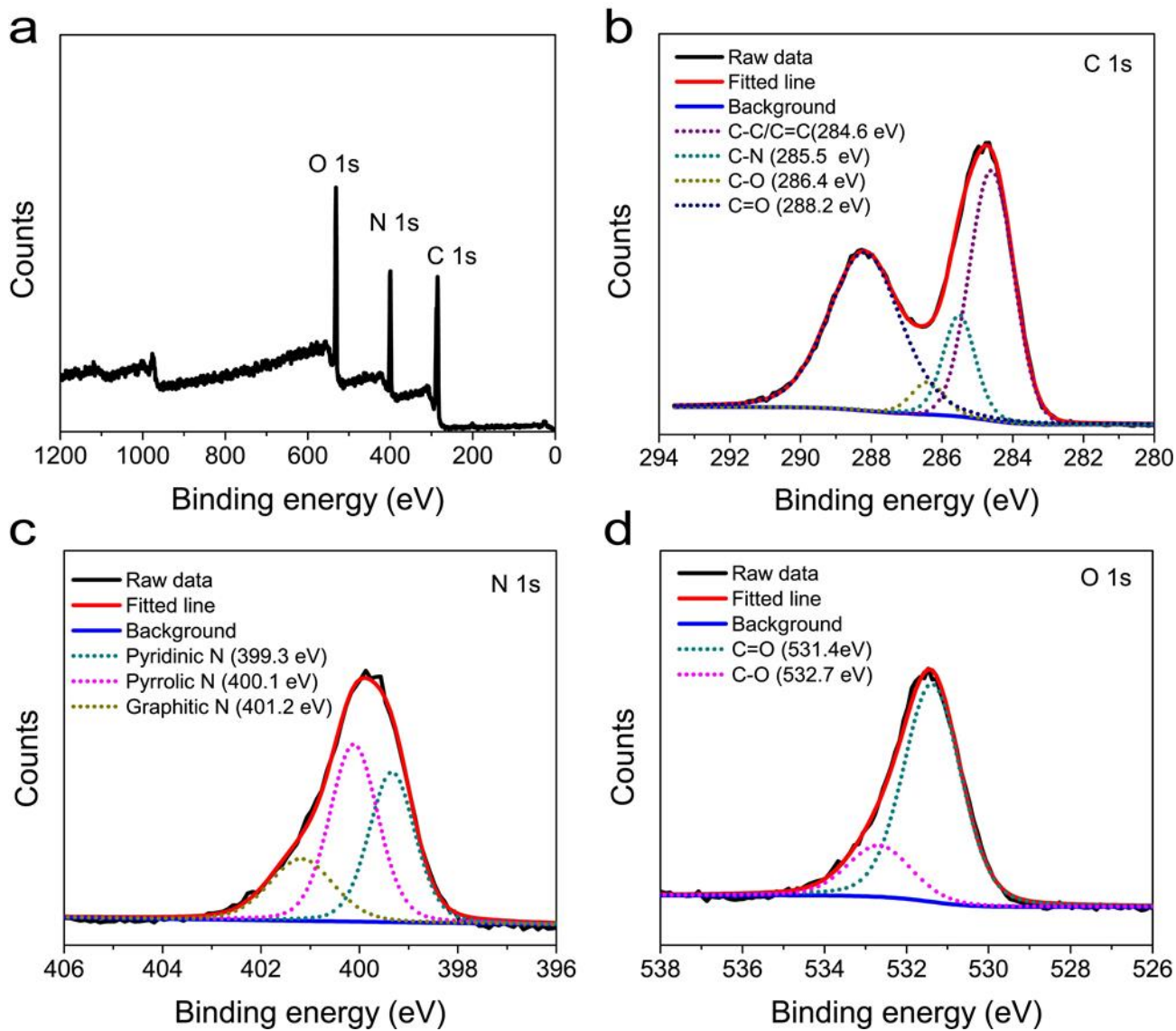

**Figure 2.** XPS survey spectrum (a) and high-resolution C 1s (b), N 1s (c), and O 1s (d) spectra of the fabricated carbon dots.

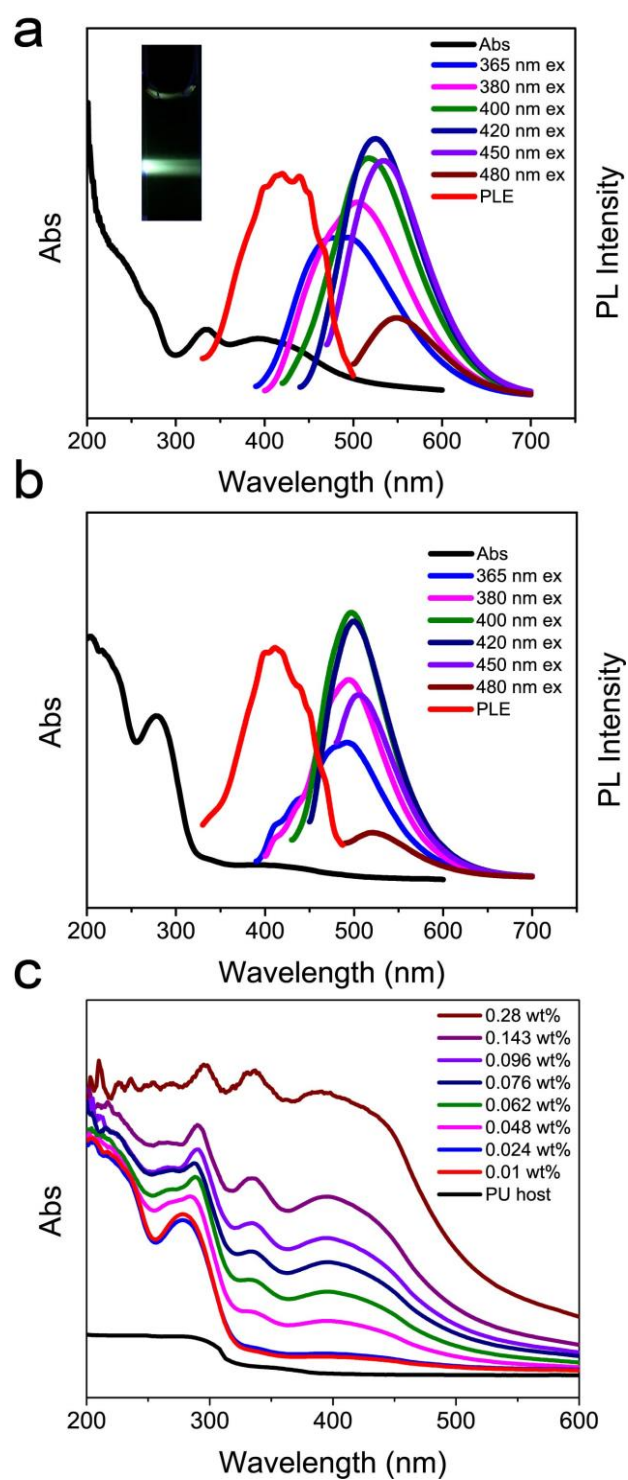

**Figure 3.** UV-Vis, PLE and PL spectra of carbon dot aqueous solution (0.01 wt%) (inset shows the photograph of CD aqueous solution under 420 nm) (a) and the PU film with 0.01 wt% carbon dots (b). (c) The UV-Vis spectra of the PU film with different carbon dot concentration.

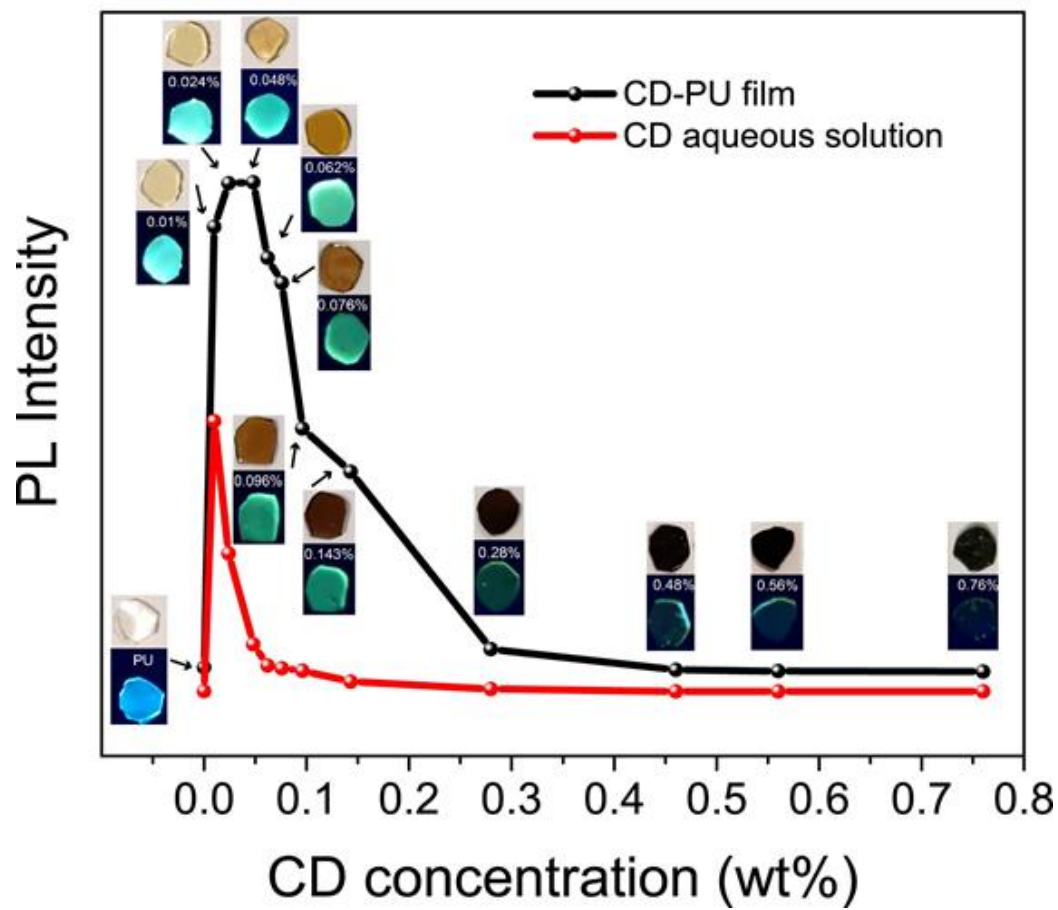

**Figure 4.** PL spectra of PU films and aqueous solutions with different carbon dot concentration. The inset shows the corresponding photographs of CD-PU film under daylight and UV light of 365 nm.

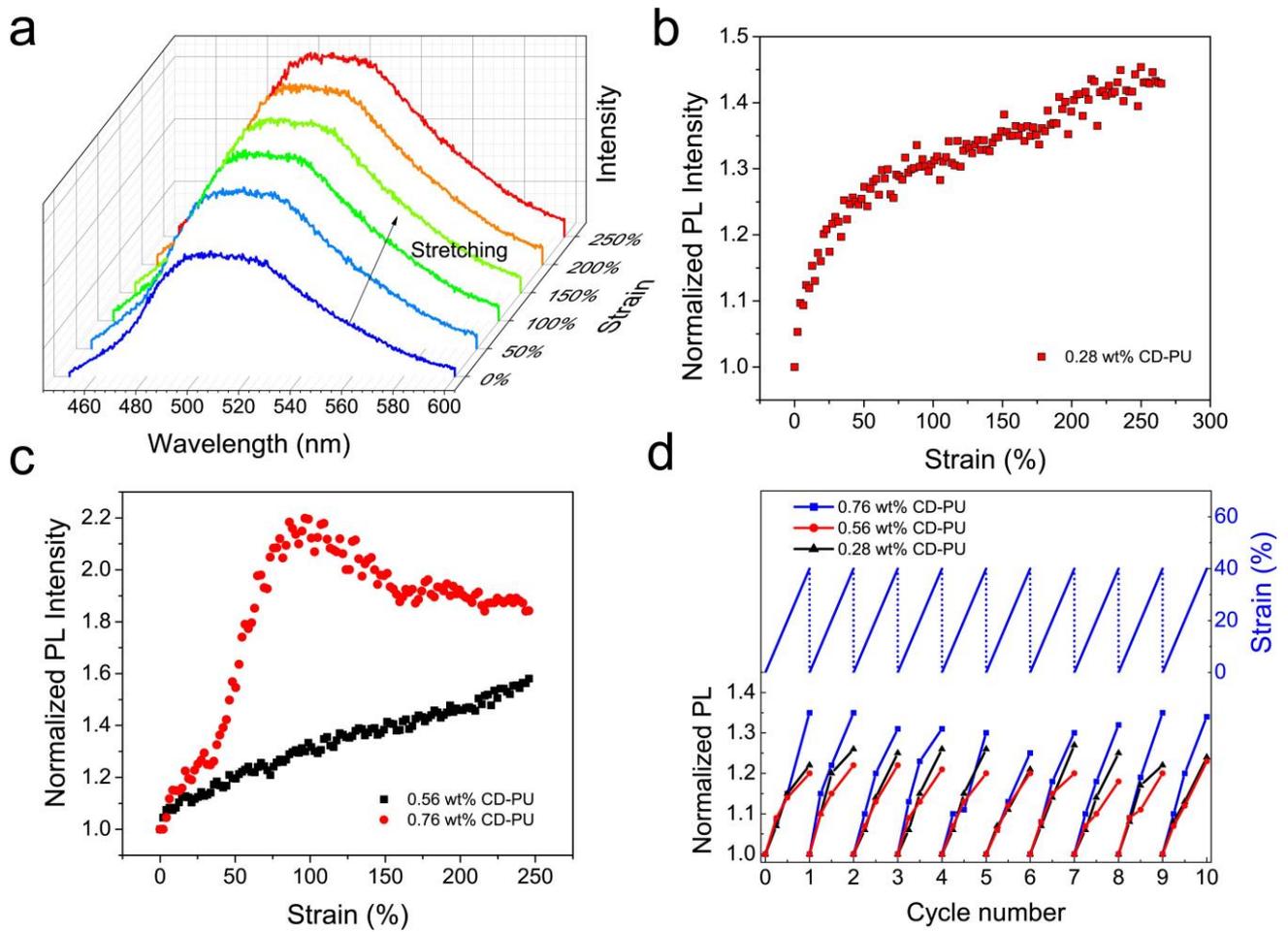

**Figure 5.** PL spectra (a) and normalized PL intensity (b) of the PU film with 0.28 wt% CDs as a function of applied uniaxial strain. (c) Normalized PL intensity of PU films with 0.56 and 0.76 wt% CDs. (d) PL intensity of the PU films with 0.28, 0.56, and 0.76 wt% CDs under repetitive strain loading.

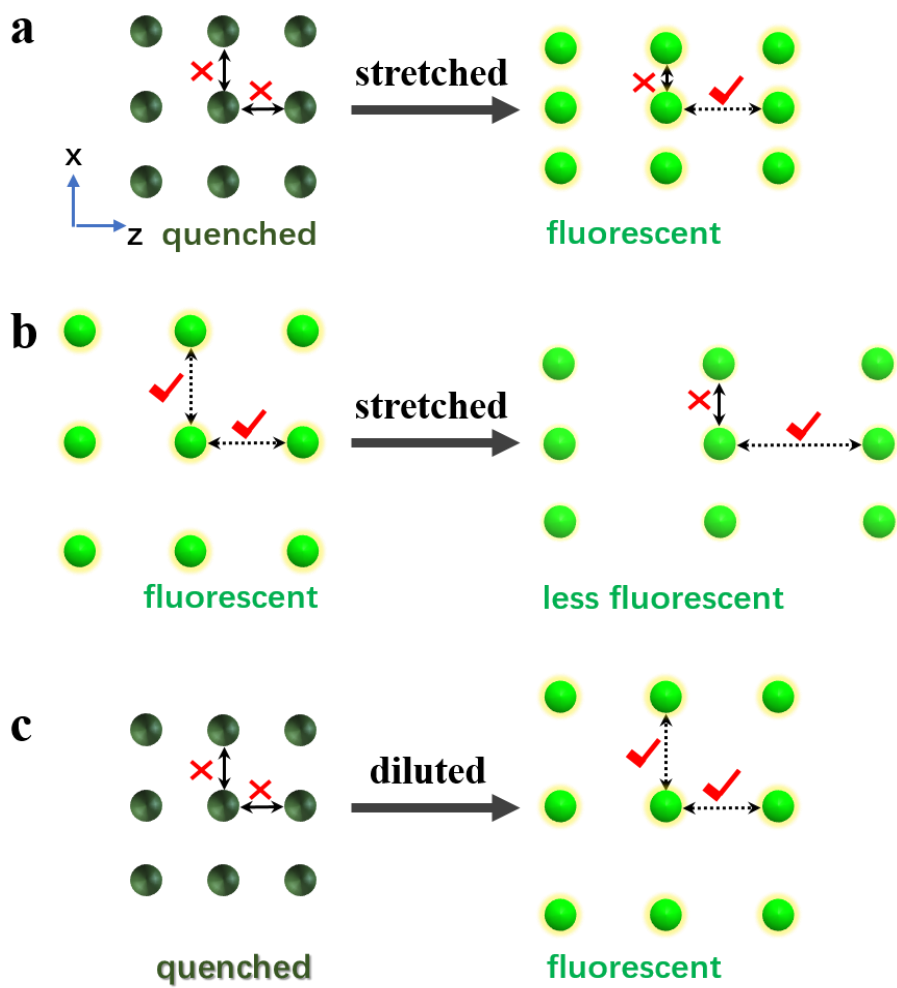

**Figure 6.** Mechanistic illustration to explain the PL responses to tensile strains for CD-PU composites with the inter-particle distance below (a) and above (b) the quenching threshold. The anti-quenching effect induced by dilution for liquid solution systems is depicted in (c) for comparison.

**Table 1.** Comparison with previously reported strain sensors

| Output signal | Transducer | Substrate | Max. Strain | Repeatability | Ref. |
|---|---|---|---|---|---|
| Optical | CD | PU | 250% | > 10 cycles | This work |
| Optical | Au-Sr$_3$Al$_2$O$_6$: Eu$^{3+}$ | Poly(dimethylsiloxane) | 18% | - | 18 |
|  | Mechanophores | Epoxy | 50% | - | 16 |
|  | Mechanophores | Nanofiber/poly(dimethylsiloxane) | 50% | 5 cycles | 17 |
|  | CdSe-CdS-tQD | Poly(styrene-ethylene-butylene-styrene) | 60% | > 20 cycles | 20 |
|  | CD | Polyethylene terephthalate | 130% | > 50 cycles | 19 |
| Electrical | rGO | Tape | 82% | > 5000 cycles | 15 |
|  | AgNW | Poly(dimethylsiloxane) | 70% | >225 cycles | 13 |
|  | SWCNT | Poly(dimethylsiloxane) | 280% | 10000 cycles | 14 |

**Table of Contents**

Flexible carbon dots-polyurethane composites are conveniently fabricated and used as low-cost nontoxic fluorescent strain sensors for detecting a wide range of strain. The sensing mechanism is attributed to a novel anisotropic anti-quenching mechanism.

**Keyword**: strain sensor

Yayuan Tian, Yan Zhao, Fengwen Kang, Fucong Lyu, Zebiao Li, Jian Lu*, and Yang Yang Li*

**Title**: **Green low-cost carbon nanodots-polyurethane composites with novel anisotropic anti-quenching mechanism for strain sensing**

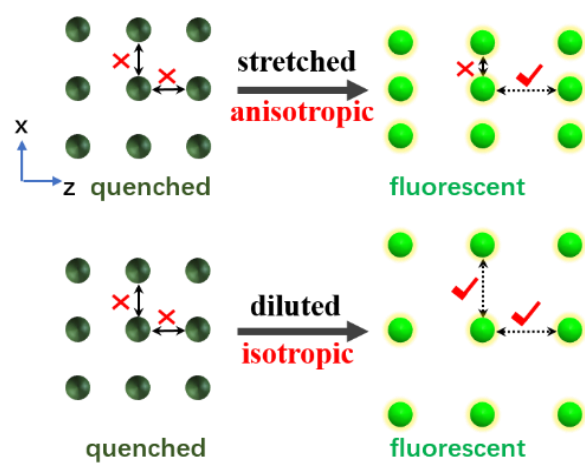

Supporting Information

**Green low-cost carbon nanodots-polyurethane composites with novel anisotropic anti-quenching mechanism for strain sensing**

Yayuan Tian, Yan Zhao, Fengwen Kang, Fucong Lyu, Zebiao Li, Jian Lu*, and Yang Yang Li*

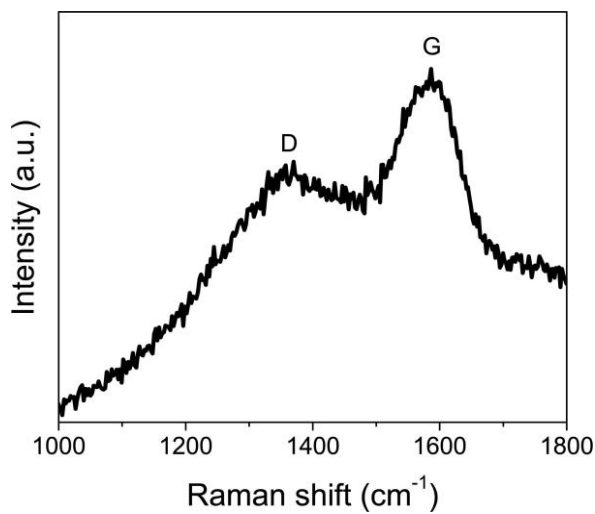

**Figure S1**. Raman spectrum of carbon nanodots.

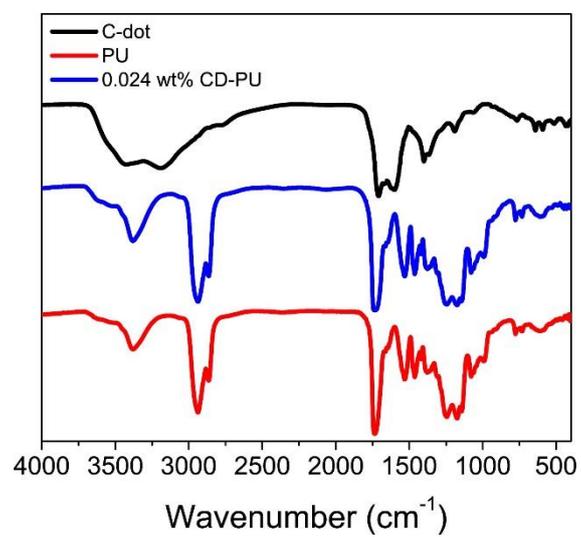

**Figure S2**. FTIR spectra of CD, PU, and CD-PU composite (0.024 wt% CD).

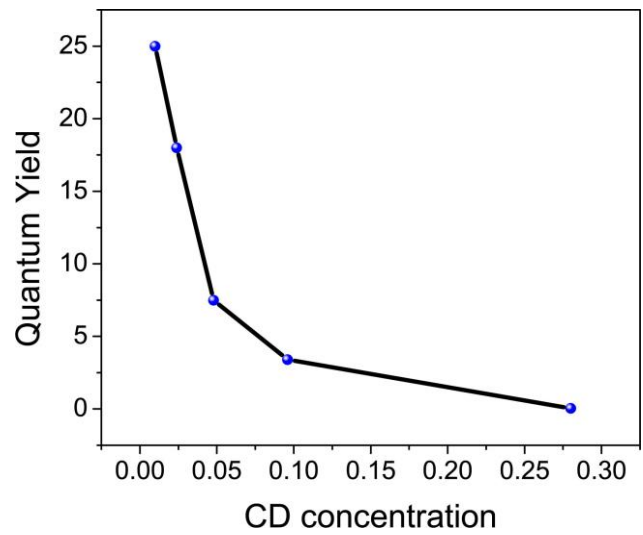

**Figure S3**. Quantum yield of CD-PU films with different CD concentration, excited at 412 nm.

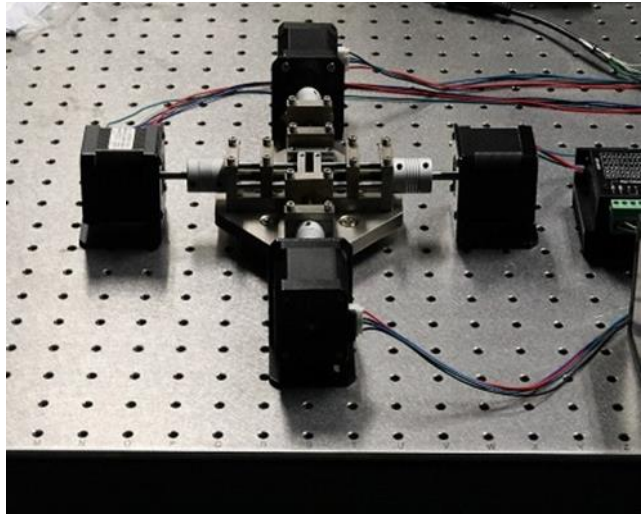

**Figure S4**. Photograph of the custom-made tension machine electrically driven by motors.

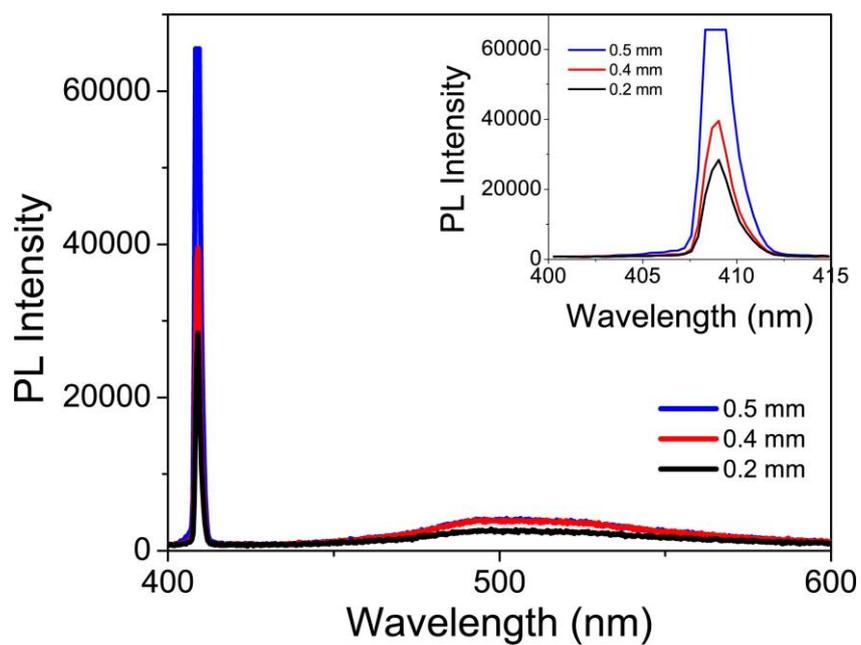

**Figure S5**. PL spectra of 0.01 wt% CD-PU film recorded with different film thickness. The inset shows an enlarged view of the excitation light intensity.

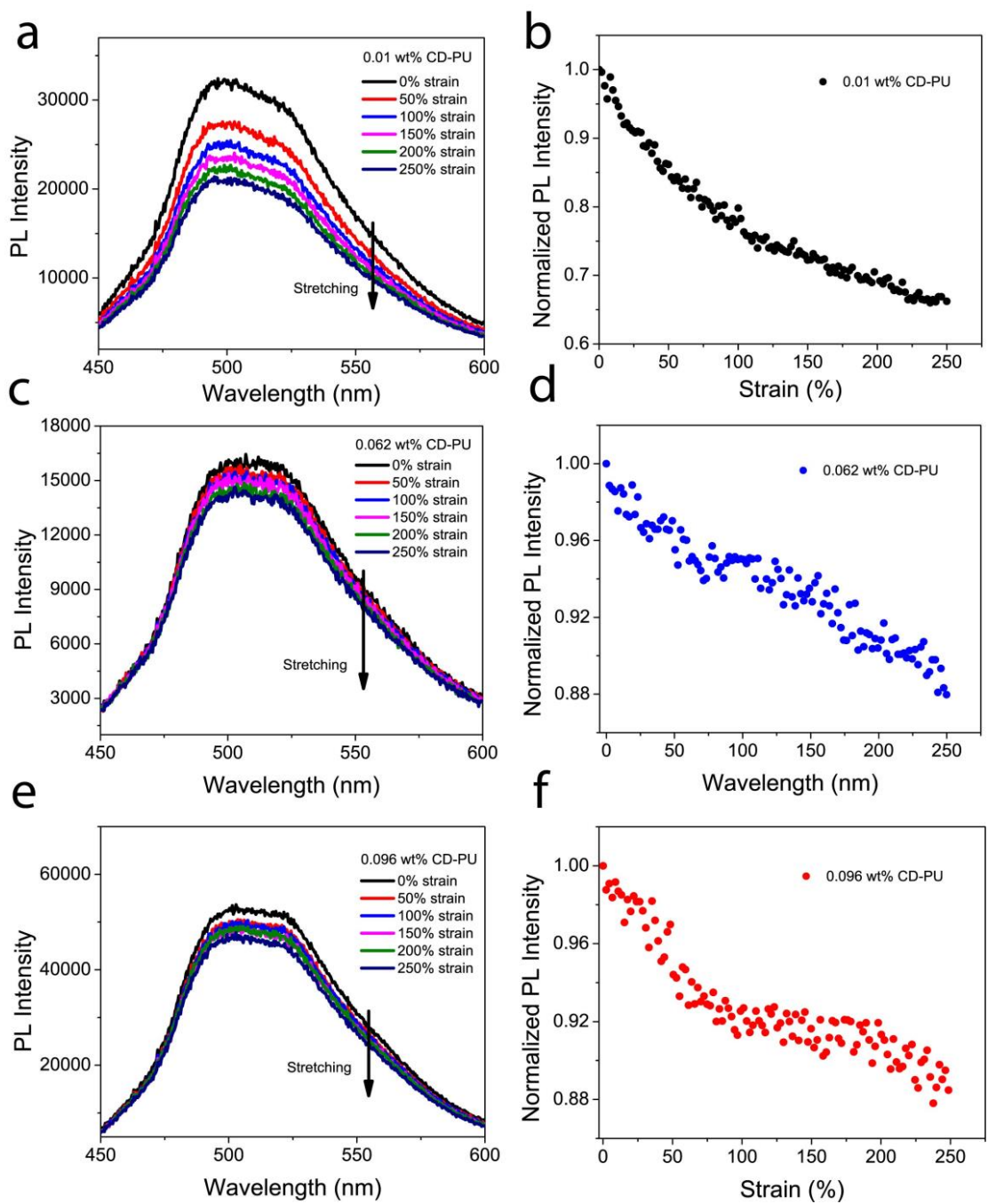

**Figure S6**. PL spectra of the CD-PU film recorded under uniaxial strain for (a) 0.01 wt% CD-PU, (c) 0.062 wt% CD-PU and (e) 0.096 wt% CD-PU. Normalized PL intensity as a function of the applied tensile strain for (b) 0.01 wt% CD-PU, (d) 0.062 wt% CD-PU and (f) 0.096 wt% CD-PU film.